\documentclass[pre,aps,floatfix,twocolumn,showpacs,amsmath,amssymb]{revtex4}

\usepackage[dvips]{graphicx}
\usepackage{dcolumn}
\usepackage{bm}

\begin{document}

\preprint{}

\title{Systematic analysis of group identification in stock markets}

\author{Dong-Hee Kim}
\email[Email address: ]{dhkim@stat.kaist.ac.kr}
\author{Hawoong Jeong}
\email[Email address: ]{hjeong@kaist.ac.kr}
\affiliation{Department of Physics,
Korea Advanced Institute of Science and Technology, Daejeon 305-701, Korea}

\date{\today}

\begin{abstract}
We propose improved methods to identify stock groups using
the correlation matrix of stock price changes.
By filtering out the marketwide effect and the random noise,
we construct the correlation matrix of stock groups in which
nontrivial high correlations between stocks are found.
Using the filtered correlation matrix, we successfully identify
the multiple stock groups without any extra knowledge of the stocks
by the optimization of the matrix representation and the percolation
approach to the correlation-based network of stocks.
These methods drastically reduce the
ambiguities while finding stock groups using the eigenvectors of the
correlation matrix.
\end{abstract}

\pacs{89.65.Gh, 05.40.Ca, 89.75.Fb, 89.75.-k }


\maketitle

\section{Introduction}
\label{sec:intro}

The study of correlations in stock markets has attracted much
interest of physicists because of its challenging complexity as
a complex system and its possible future applications
to the real markets~\cite{book}. In the early years, a correlation-based
taxonomy of stocks and stock market indices was studied
by the method of the hierarchical tree~\cite{Mantegna,Bonanno1}.
Recently, the minimum spanning tree technique was introduced to
study the structure and dynamics of the stock network
\cite{Bonanno2,Onnela1,Onnela2},
the random matrix theory was applied to find out the difference between
the random and nonrandom property of the correlations
\cite{Laloux,Plerou1,Plerou2,Gopikrishnan,Utsugi}, and
the maximum likelihood clustering method was developed and
applied to identify cluster structures in stock markets~\cite{Giada}.
Also, these studies have been extended to the applications to
the portfolio optimization in real market~\cite{Plerou2,Onnela1}.

Commonly, the correlation between stocks is expressed by the Pearson
correlation coefficient of log-returns,
\begin{equation}
\label{eq:log_return}
G_i(t) \equiv \ln S_i(t+\Delta t) - \ln S_i(t),
\end{equation}
where $S_i(t)$ is the price of stock $i$ at time $t$.
From real time series data
of $N$ stock prices, we can calculate the element of $N \times N$ correlation
matrix $\mathbf{C}$ as following
\begin{equation}
\label{eq:cij}
C_{ij} = \frac{ \langle (G_i(t)-\langle G_i \rangle)
(G_j(t)-\langle G_j \rangle) \rangle}
{\sqrt{(\langle G_i^2 \rangle - \langle G_i \rangle^2)
(\langle G_j^2 \rangle - \langle G_j \rangle^2)}} ,
\end{equation}
where $\langle \cdots \rangle$ indicates time
averages over the period of the time series. By definition,
$C_{ii}=1$ and $C_{ij}$ has a value in $[-1,1]$.

Laloux {\it et al.}~\cite{Laloux} and
Plerou {\it et al.}~\cite{Plerou1,Plerou2} studied the statistical
properties of an empirical correlation matrix between stock price changes
defined in Eq.~(\ref{eq:cij}) for real markets. In comparison with
the prediction of the random matrix theory, they found that
the statistics of the bulk eigenvalues are in remarkable agreements
with the universal properties of the random correlation matrix.
For example, the bulk part of the eigenvalue
spectrum of the empirical correlation matrix for $N$ stocks over $L$
price data has the form of the spectrum of the random correlation
matrix~\cite{Sengupta} which is given by
\begin{equation}
\label{eq:spectrum}
\rho(\lambda) = \frac{Q}{2\pi}
\frac{\sqrt{(\lambda_{max} - \lambda)(\lambda - \lambda_{min})}}{\lambda} ,
\end{equation}
for $\lambda \in [\lambda_{min},\lambda_{max}]$ in the limit of
$N,L \to \infty$ with fixed $Q \equiv L/N$, where
$\lambda_{max}=(1+1/\sqrt{Q})^2$ and $\lambda_{min}=(1-1/\sqrt{Q})^2$.
Moreover, the level spacing statistics of eigenvalues exhibits
good agreement with the results from the Gaussian orthogonal ensemble
of random matrices \cite{Plerou1,Plerou2}.

On the other hand,
the nonrandom properties of the correlation matrix have also been studied
with the empirical correlation matrix~\cite{Plerou1,Plerou2,Gopikrishnan}.
From the empirical data for the New York Stock Exchange,
it was found that each eigenvector corresponding to the few largest eigenvalues
larger than the upper bound of the bulk eigenvalue spectrum, is
$\it localized$, in a sense that only a few components contribute
to the eigenvector mostly, and the stocks corresponding to those dominant
components of the eigenvector are found to belong to a common industry sector.
Very recently, Utsugi {\it et al.} confirmed and
improved those results through the similar analysis for the Tokyo Stock
Exchange~\cite{Utsugi}.

\begin{figure}
\begin{center}
\includegraphics[width=0.4\textwidth]{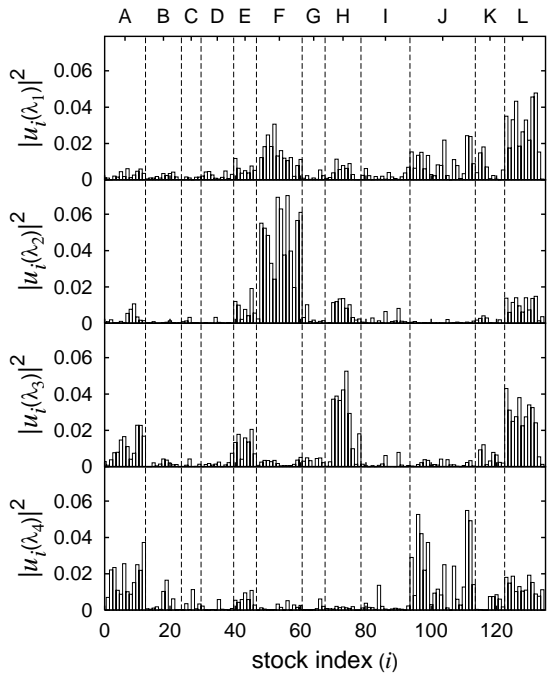}
\caption{
\label{fig:eigenvector}
The normalized eigenvector components $u_i(\lambda)$ of stock $i$
corresponding to the second to fifth largest eigenvalues 
$\lambda_1$ - $\lambda_4$
of the correlation matrix. The stocks are sorted by industrial sectors,
A: basic materials, B: capital goods, C: conglomerates, D: consumer (cyclical),
E: consumer (noncyclical), F: energy, G: financial, H: healthcare,
I: services, J: technology, K: transportation, and L: utilities,
which are separated by dashed lines.
}
\end{center}
\end{figure}

In order to confirm the localization property of eigenvectors,
we perform the similar analysis to the previous studies
\cite{Plerou1,Plerou2,Gopikrishnan} on eigenvectors of
the correlation matrix using our own dataset of stock prices.
We analyze the daily prices of $N=135$ stocks belonging to the New York
Stock Exchange (NYSE) for the 20-year period $1983-2003$
($L\simeq 5000$ trading days) which is publicly available from the
web-site(http://finance.yahoo.com)~\cite{comment0}.
Indeed, if we put stocks in the order of their industrial sectors,
we observe that the eigenvector components corresponding to stocks
which belong to specific industrial sectors
give high contributions to each of the eigenvectors for
the few largest eigenvalues (see Fig.~\ref{fig:eigenvector}).
For instance, the stocks belonging to the energy, technology,
transportation, and utilities sectors highly contribute to the eigenvector
for the second largest eigenvalue; the energy sector constitutes the big
part of the eigenvector for the third largest eigenvalue; the fourth
largest eigenvalue gives the eigenvector localized on the basic materials,
consumer (noncyclical), healthcare, and utilities sectors; the eigenvector
for the fifth largest eigenvalue is also localized on several specific
industrial sectors.

However, it is not straightforward to find out specific stock groups, such as
the industrial sectors, inversely. If each of the eigenvectors had
well-defined dominant components and the corresponding set of stocks
were independent of the sets from other eigenvectors, it would become
easy to identify the stock groups. Unfortunately, in our study,
it turns out that
not only the set of eigenvector components with dominant contribution
can be hardly defined in the eigenvector but also such a set
is likely to overlap with the sets from other
eigenvectors unless we pick a very small number of stocks with
few highest ranks of their contributions to the eigenvectors;
Figure~\ref{fig:eigenvector} indicates that each of the eigenvectors
is localized on a multiple number of industrial sectors and the
corresponding stocks severely overlap with those from the other eigenvectors.
Therefore it is very ambiguous to identify the stock groups
for practical purposes. The aim of this study is to get rid of these
ambiguities and finally find out relevant stock groups
without any aid of the table of industrial sectors.

In this paper, we introduce the improved method to identify
stock groups which drastically reduce the ambiguities in finding
multiple groups using eigenvectors of the correlation matrix.
We first filter out the random noise
and the marketwide effect from the correlation matrix. With
the filtered correlation matrix, we apply optimization and
percolation approaches to find the stock groups.
Through the optimization of the stock sequences
representing the matrix indices,
the filtered correlation matrix is transformed into
the block diagonal matrix in which all stocks
in a block are found to belong to the same group.
By constructing a network of stocks using the percolation approach
on the filtered correlation matrix,
we also successfully identify the stock groups which appear in the form
of isolated clusters in the resulting network.

This paper is organized as follows. In Sec.~\ref{sec:dmatrix},
the detailed filtering method to construct the group correlation
matrix is given.
For the filtering, the largest eigenvalue and the corresponding
eigenvector are required and they are calculated from
the first-order perturbation theory.
In Sec.~\ref{sec:group}, detailed stock group finding methods
using the optimization and the percolation are given and the resulting
stock groups are specified. In Sec.~\ref{sec:conclusion},
a summary and conclusions are presented.

\section{Group correlation matrix}
\label{sec:dmatrix}

\subsection{Filtering}
\label{subsec:filtering}

The group of stocks is defined as a set of highly intercorrelated
stocks in their price changes. In the
empirical correlation matrix, because several types of noises are
expected to coexist with the intragroup correlations, it is
essential to filter out such noises to isolate the intragroup
correlations which we are interested in.
With the complete set of eigenvalues and eigenvectors, the correlation
matrix in Eq.~(\ref{eq:cij}) can be expanded as
\begin{equation}
\mathbf{C}=\sum^{N-1}_{\alpha=0}
\lambda_\alpha | \alpha \rangle \langle \alpha |,
\end{equation}
where $\lambda_\alpha$ is the eigenvalue sorted in descending order
and $|\alpha \rangle$ is the corresponding eigenvector.
Because only the eigenvectors corresponding to the few largest eigenvalues
are believed to contain the information on significant stock groups, we can
identify a filtered correlation matrix for stock groups by
choosing a partial sum of $\lambda_\alpha | \alpha \rangle \langle \alpha |$
relevant to stock groups,
which we will call {\it the group correlation matrix}, $\mathbf{C}^g$.

In order to extract $\mathbf{C}^g$ from the correlation matrix,
taking the previous results of Plerou {\it et al.}
~\cite{Gopikrishnan,Plerou1,Plerou2} for granted,
we posit that the eigenvalue spectrum of the correlation
matrix is organized by the marketwide part of the largest eigenvalue,
the group part of intermediate discrete eigenvalues, and the random part
of small bulk eigenvalues.
Then, we can separate the correlation matrix into three parts as
\begin{eqnarray}
\label{eq:filter}
\mathbf{C} &=& \mathbf{C}^m + \mathbf{C}^g + \mathbf{C}^r \nonumber \\
&=& \lambda_0 | 0 \rangle \langle 0 | +
\sum^{N_g}_{\alpha =1} \lambda_\alpha | \alpha \rangle \langle \alpha | +
\sum^{N-1}_{\alpha =N_g+1} \lambda_\alpha | \alpha \rangle \langle \alpha | ,
\end{eqnarray}
where $\mathbf{C}^m$, $\mathbf{C}^g$, and $\mathbf{C}^r$ indicate
the marketwide effect, the group correlation matrix, and the random noise
terms, respectively.

\begin{figure}
\begin{center}
\includegraphics[width=0.4\textwidth]{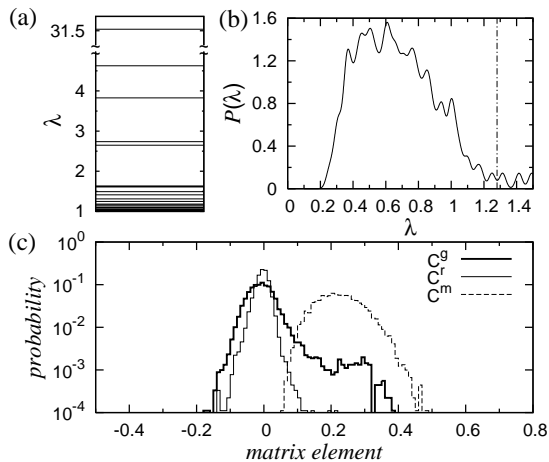}
\caption{
\label{fig:distribution}
(a) The eigenvalues $\lambda > 1.0$ of the correlation matrix $\mathbf{C}$ and
(b) the distribution of bulk eigenvalues $P(\lambda )$ (solid line).
The dashed-dot line marks our boundary between the random noise part and
the group correlation part.
(c) The matrix element distribution
for the group correlation matrix $\mathbf{C}^g$ and
the residual parts corresponding to the bulk eigenvalues
$\mathbf{C}^r$ and the largest eigenvalue $\mathbf{C}^m$.
}
\end{center}
\end{figure}

While the determination of $\mathbf{C}^m$ is straightforward,
it is not so clear to determine $N_g$ for separating $\mathbf{C}^g$ and
$\mathbf{C}^r$. If there were no correlation between stock prices,
the bulk eigenvalues have to follow Eq.~(\ref{eq:spectrum}), and thus
the upper bound of the bulk eigenvalues can be clearly determined
from $Q$. However, in empirical correlation matrix, the bulk eigenvalue
spectrum deviates from Eq.~(\ref{eq:spectrum}) due to the coupling with
underlying structured correlations, such as the group correlation embedded in
$\mathbf{C}^g$ \cite{Burda}.
Therefore we use a graphical estimation to determine $N_g$;
in the eigenvalue spectrum as shown in Fig.~\ref{fig:distribution}(b)
we choose the cut $N_g=9$ in the vicinity of the blurred tail
of the bulk part of the spectrum. Nevertheless, in spite of
the rough estimation of $N_g$, we note that our results in
this work do not alter from a small change of $N_g$, $\sim \pm 1$.
This can be justified by the following arguments. In the group correlation 
matrix, the corresponding component of the eigenvalues close to 
the bulk part of the spectrum is confined to only a very small portion 
of the whole matrix; because the elements of the correlation matrix 
component $\lambda_\alpha | \alpha \rangle \langle \alpha |$ must be smaller 
than the eigenvalue $\lambda_\alpha$, large discrete eigenvalues 
dominantly contribute to the group correlation matrix. 
In addition, even if we count one less eigenvalue near the boundary 
of bulk part of the spectrum in constructing the group correlation matrix, 
a possible information loss of groups is not likely serious 
because the pure eigenvectors of the groups generally turn out 
to be mixed all together in the eigenvectors of the correlation matrix 
(see Fig.~\ref{fig:eigenvector}). 
Therefore the influence from the error in the determination 
of $N_g$ is insignificant so that it does not change the clustering result.

This decomposition of the correlation matrix gives nontrivial
characteristics to the distribution of the group correlation matrix elements
$C^g_{ij}$. In Fig.~\ref{fig:distribution}(c), it turns out that
the distribution of $C^g_{ij}$ shows positive heavy tail. This indicates
that $\mathbf{C}^g$ contains a non-negligible number of
strongly correlated stock pairs, which is expected to come from the
correlation between the stocks belonging to the same group.
On the other hand, $\mathbf{C}^r$ shows the Gaussian distribution
consistent with the prediction of the random matrix theory~\cite{Plerou2}.
While this Gaussian-like distribution is also observed partially
in the distribution of $C^g_{ij}$ due to the coupling between
group correlations and random noises, it turns out 
that this remaining noise does not seriously affect the 
identification of stock groups.
The distribution of $C^m_{ij}$ shows that $\mathbf{C}^m$ also contains
highly correlated stock pairs, but we find that $C^m_{ij}$  is
not relevant to the group correlation and thus have to be filtered out
for the clear identification of  the stock groups,
which is discussed in Sec.~\ref{subsec:largest}.

\begin{figure}
\begin{center}
\includegraphics[width=0.4\textwidth]{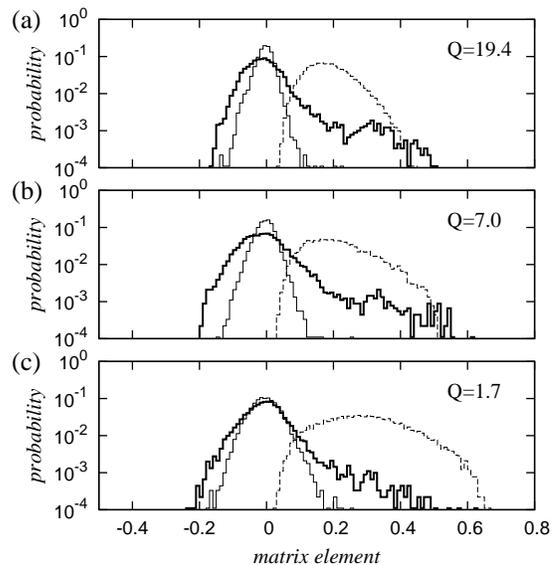}
\caption{
\label{fig:Qdep}
The $Q \equiv L/N$ dependence of the matrix element distribution for 
$\mathbf{C}^g$ (thick solid line), $\mathbf{C}^r$ (solid line), and
$\mathbf{C}^m$ (dashed line). With fixed $N=135$, various time periods are 
tested for (a) $L \simeq 2600$ ($1993-2003$), (b) $L \simeq 950$ ($2000-2003$), 
and (c) $L \simeq 240$ ($2003$).
}
\end{center}
\end{figure}

Since the quality of the correlation matrix can depend on
the period of empirical data or generally $Q \equiv L/N$,
our decomposition of the correlation matrix can also depend on
$Q$. Here we simply check how the determination of $N_g$ and
the resulting matrix element distribution of the decomposed matrices
are changed depending on $Q$ (see Fig.~\ref{fig:Qdep}). 
For $Q=19.4$ ($1993-2003$) and
$Q=7.0$ ($1999-2003$), $\mathrm{C}^g$ and $\mathrm{C}^r$ are
separated at $N_g=10$, which is not very different from $N_g=9$
of the larger dataset we use throughout this paper, and
in addition, the distribution of the matrix element shows
the similar degree of the heavy tail in $C^g_{ij}$.
However, decreasing $Q$ much smaller, the bulk eigenvalue spectrum
becomes wider so that more eigenvalues relevant to the group
correlation can be buried in the bulk spectrum, which
leads to smaller $N_g$ that turns to be $7$ for $Q=1.7$ (2003).
Even in this case of $Q=1.7$, the positive heavy tail is still found
in $C^g_{ij}$ but very weaker than higher $Q$'s.
These imply that we need a large enough $Q$ for the stock group
identification.

\subsection{Largest eigenvalue and corresponding eigenvector}
\label{subsec:largest}

Our filtering is based on the following interpretations
of the previous studies:
the bulk part of the eigenvalues and their
eigenvectors are expected to show the universal properties of the random
matrix theory and the largest eigenvalue and its eigenvector are considered
as a collective response of the entire market
\cite{Plerou1,Plerou2,Gopikrishnan}.
While the random characteristics of the bulk eigenvalues
have been studied intensively, only the empirical tests have been done
for the largest eigenvalue and its eigenvector so far
\cite{Plerou2,Gopikrishnan}.
Thus, to understand the more accurate meaning,
we calculate the largest eigenvalue and
its eigenvector of the correlation matrix by using perturbation theory.

In stock markets, it has been understood that there exist three kinds of
fluctuations in stock price changes: a marketwide fluctuation,
synchronized fluctuations of stock groups, and a random fluctuation
\cite{Plerou1,Plerou2,Gopikrishnan}.
For simplicity, we consider a situation
in which a system with only the marketwide fluctuation
is perturbed by other fluctuations.
Let us assume that the price changes of all the stocks
in the market find a synchronized background fluctuation
with zero mean and variance $c_0$ as a marketwide effect.
Then, we can write down
the $N \times N$ unperturbed correlation matrix as
\begin{equation}
\mathbf{C}^0=
\left( \begin{array}{cccc}
1 & c_0 & \cdots & c_0 \\
c_0 & 1 & & \vdots \\
\vdots & & \ddots & c_0 \\
c_0 & \cdots & c_0 & 1
\end{array} \right) ,
\end{equation}
which has the largest eigenvalue
$\lambda^{(0)}_0 = c_0 (N-1) +1$ and its eigenvector components
$u^{(0)}_i = \langle (stock)~i | 0^{(0)} \rangle = 1/\sqrt{N}$.

When a small perturbation is turned on, the total correlation matrix
becomes
\begin{equation}
\label{eq:perturbC}
\mathbf{C} = \mathbf{C}^0 + \mathbf{\Delta} ,
\end{equation}
where $\Delta_{ii} = 0 $ and $\Delta_{ij}=\Delta_{ji}$.
Applying the perturbation theory up to the first order,
the largest eigenvalue and the corresponding eigenvector components
are easily calculated as
\begin{eqnarray}
\label{eq:eigen}
\lambda_0 &=& c_0 (N-1) + 1 + \frac{1}{N}\sum_{i,j}\Delta_{ij}, \nonumber \\
u_i &=& \frac{1}{c_0 N^{3/2}}
\left( w_i + c_0 - \frac{1}{N}\sum_{j,k}\Delta_{jk}
\right),
\end{eqnarray}
where $w_i = \sum_{j \neq i} C_{ij}$.

We check the validity of Eqs.~(\ref{eq:eigen}) by
comparing with the largest eigenvector
obtained from the numerical diagonalization
of the empirical correlation matrix. For the comparison, we make
the distribution of $C_{ij}$ in Eq.~({\ref{eq:perturbC})
to be close to the empirical $C_{ij}$ distribution by
assuming that $\Delta_{ij}$ follows the bell-shaped
distribution with zero mean and letting $c_0$ to the mean value
of the empirical $C_{ij}$. Because the assumption not only reproduces
the distribution of empirical $C_{ij}$, but also allows us to neglect
the $1/N \sum \Delta_{ij}$ term in Eqs.~({\ref{eq:eigen}), we
can directly compare the perturbation theory with the numerical result.
Figure~\ref{fig:compare} displays the eigenvector components of the
largest eigenvalue obtained from the empirical correlation matrix and
the dominant terms of Eqs.~(\ref{eq:eigen}), which show remarkable agreement
with each other.

\begin{figure}
\begin{center}
\includegraphics[width=0.4\textwidth]{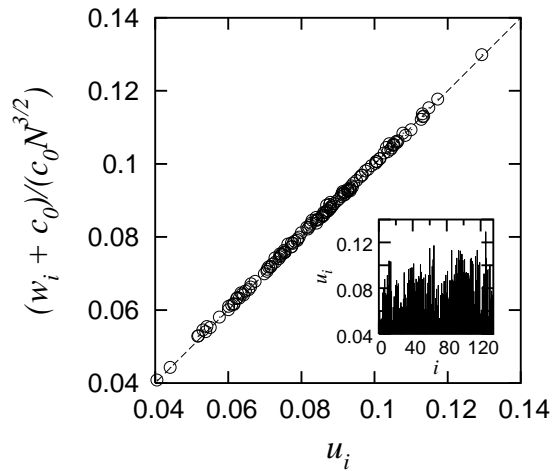}
\caption{
\label{fig:compare}
The comparison of the eigenvector of the largest eigenvalue
obtained by the exact diagonalization $u_i$ and the dominant term
$(w_i + c_0)/(c_0 N^{3/2})$ in Eq.~(\ref{eq:eigen}). The dashed line
has the slope $1.0$.
(Inset: the values of corresponding eigenvector components.)
}
\end{center}
\end{figure}

Equation~(\ref{eq:eigen}) indicates that the eigenvector of
the largest eigenvalue is contributed by not only the global fluctuation
but also the unknown perturbations from $\mathbf{\Delta}$
including random noises. Thus, by filtering out the $\mathbf{C}^m$
term, we can decrease the effect of unnecessary perturbations
in constructing the group correlation matrix.
Indeed, as seen in Fig.~\ref{fig:distribution}(c),
because the heavy tail part of $\mathbf{C}^g$, the highly correlated elements,
are buried in $\mathbf{C}^m$, the clustering of stocks would be
seriously disturbed unless $\mathbf{C}^m$ is filtered out.

In addition, Eqs.~(\ref{eq:eigen}) also enable us to interpret
more detailed meaning of the eigenvector than the marketwide effect.
Because the $i$th eigenvector component $u_i$ is mostly determined by $w_i$,
the sum of the correlation over all the other stocks, it can be regarded as
\textit{the influencing power} of the company in the entire stock market.
In real data, the top four stocks with highest $w_i$ are found to be
General Electric (GE), American Express (AXP),
Merrill Lynch (MER), and Emerson Electric (EMR),
mostly conglomerates or huge financial companies, which convinces us
that $u_i$ is indeed representing the influencing power of stock $i$.
However, these high influencing companies prevent clear clustering of
stocks because of their non-negligible correlations with entire stocks
in the market. This is easily comprehensible by considering
an analogous situation in a network where the big hub, a node with
a large number of links, can make indispensable
connections between groups of nodes to cause difficulties in distinguishing
the groups~\cite{Holme}.
Therefore it is very important to filter out $\mathbf{C}^m$
in order to identify the groups of stocks efficiently.

\section{Identification of stock groups}
\label{sec:group}

In the group model for stock price correlation proposed by Noh~\cite{Noh},
the correlation matrix $\mathbf{C}$ takes the form of
$\mathbf{C}=\mathbf{C}^g+\mathbf{C}^r$, where $\mathbf{C}^g$ and
$\mathbf{C}^r$ are the correlation matrix of stock groups and
random correlation matrix, respectively. The model assumes the ideal
situation with $C^g_{ij}=\delta_{\alpha_i,\alpha_j}$, where $\alpha_i$
indicates the group to which the stock $i$ belongs. Thus $\mathbf{C}^g$
is the block diagonal matrix,
\begin{equation}
\mathbf{C}^g=
\left( \begin{array}{cccc}
\mathbf{1}_0 & 0 & \cdots & 0 \\
0 & \mathbf{1}_1 & & \vdots \\
\vdots & & \ddots & 0 \\
0 & \cdots & 0 & \mathbf{1}_n
\end{array} \right) ,
\end{equation}
where $\mathbf{1}_i$ is the $N_i \times N_i$ matrix ($N_i$ is the number of
stocks in the $i$th group) of which all elements are $1$.

Here we use this group model to find the groups of stocks.
If the correlation matrix in the real market can be represented
by the block diagonal matrix as in the model, it would be very easy to
identify the groups of stocks. However, there exist infinitely
many possible representations of the matrix depending on indexing
of rows and columns even if we have a
matrix equivalent to the block diagonal matrix.
For instance, if we exchange the indices of the matrix
(e.g., $\{ i,j,k \} \to \{ k,i,j \}$)
the matrix may not be block-diagonal anymore. 
Therefore the problem in identifying
the groups in stock correlation matrix requires one
to find out the optimized sequence of stocks to transform the
matrix into the well-organized block diagonal matrix~\cite{comment1}.

\begin{figure}
\begin{center}
\includegraphics[width=0.4\textwidth]{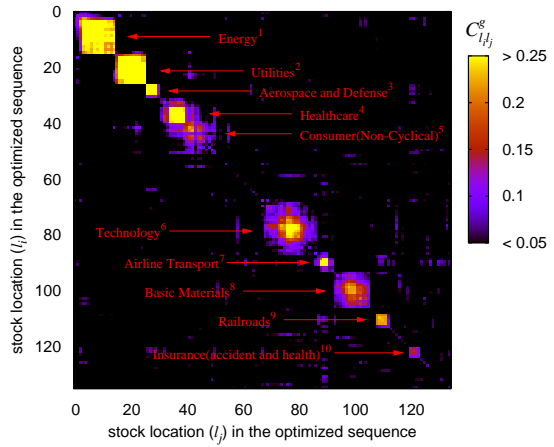}
\caption{
\label{fig:block}
(Color online) The visualization of the group correlation matrix with
the optimized stock sequence $\{l_i\}$.
}
\end{center}
\end{figure}

To optimize the sequence of stocks for clear block diagonalization,
we consider the correlation between
two stocks as an attraction force between them.
For the ideal group correlation matrix in the group model,
the block diagonal form is evidently the most stable form
if the attractive force between stocks is proportional to
their correlation within the group.
To deal with the real correlation matrix,
we define the total energy for a stock sequence as
\begin{equation}
\label{eq:energy}
E_{tot} = \sum_{i<j} C^g_{ij} |l_i-l_j| \Theta (C^g_{ij}-c_c) ,
\end{equation}
where $l_i$ is the location of the stock $i$ in the new index sequence and
the cutoff $c_c=0.1$ is introduced to get rid of the random noise
part which still remains in $\mathbf{C}^g$ in spite of the
filtering~\cite{comment2}.

\begin{figure}
\begin{center}
\includegraphics[width=0.4\textwidth]{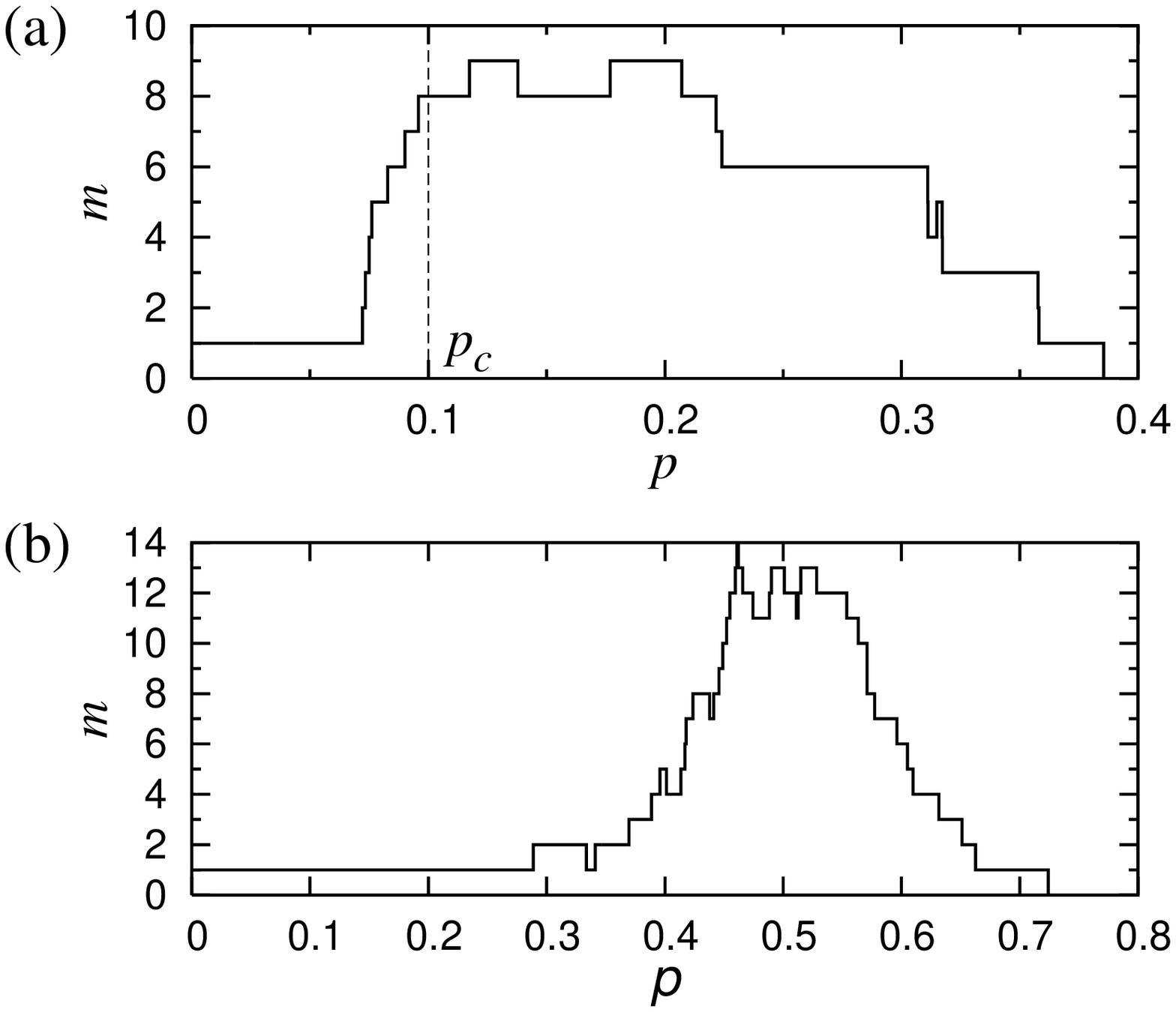}
\caption{
\label{fig:cluster}
The dependence of the number of isolated clusters, $m$, in the stock network
on the threshold $p$ in constructing the network from
(a) the group correlation matrix and (b) the full correlation matrix.
}
\end{center}
\end{figure}

We obtain the optimized sequence of stocks to minimize the total energy
defined in Eq.~(\ref{eq:energy}) by using the simulated
annealing technique~\cite{Kirkpatrick} in Monte Carlo simulation.
The following description of our problem is very similar to the well-known
{\it traveling salesman problem}, finding an optimized sequence of
visiting cities which minimizes total traveling distance~\cite{nr}:
\begin{enumerate}
\item {\it Configuration}. The stocks are numbered $i=0, \ldots , N-1$. A
configuration, a sequence of stocks $\{l_i\}$, is a permutation
of the numbers $0, \ldots , N-1$.
\item {\it Rearrangements}. A randomly chosen stock in the sequence is removed
and inserted at the random position of the sequence.
\item {\it Objective function}. We use $E_{tot}$ in Eq.~(\ref{eq:energy})
as an objective function to be minimized after rearrangements.
\end{enumerate}

\begin{table*}
\caption{\label{tab}The full list of the optimized sequence of stocks.
The footnotes correspond to the identified stock groups represented by the
same footnotes in Fig.~\ref{fig:block}.
}
\begin{ruledtabular}
\begin{tabular}{lcllcllcl}
\multicolumn{1}{c}{$p_i$} & \multicolumn{1}{c}{Ticker} & \multicolumn{1}{c}{Sector} &
\multicolumn{1}{c}{$p_i$} & \multicolumn{1}{c}{Ticker} & \multicolumn{1}{c}{Sector} &
\multicolumn{1}{c}{$p_i$} & \multicolumn{1}{c}{Ticker} & \multicolumn{1}{c}{Sector} \\
\hline
0 & XNR & Services & 45 & G & Consumer noncyclical$^5$ & 90 & AMR & Transportation$^7$ \\
1 & WMB & Utilities & 46 & AVP & Consumer noncyclical$^5$ & 91 & F & Consumer cyclical$^7$ \\
2 & VLO & Energy$^1$ & 47 & MCD & Services & 92 & GM & Consumer cyclical$^7$ \\
3 & NBL & Energy$^1$ & 48 & IFF & Basic materials & 93 & HPC & Basic materials$^8$ \\
4 & APA & Energy$^1$ & 49 & WMT & Services & 94 & DD & Basic materials$^8$ \\
5 & KMG & Energy$^1$ & 50 & FNM & Financial & 95 & CAT & Capital goods$^8$ \\
6 & HAL & Energy$^1$ & 51 & EC & Consumer cyclical & 96 & DOW & Basic materials$^8$ \\
7 & SLB & Energy$^1$ & 52 & KR & Services & 97 & WY & Basic materials$^8$ \\
8 & BP & Energy$^1$ & 53 & HET & Services & 98 & IP & Basic materials$^8$ \\
9 & COP & Energy$^1$ & 54 & TXI & Capital goods & 99 & GP & Basic materials$^8$ \\
10 & CVX & Energy$^1$ & 55 & FO & Conglomerates & 100 & BCC & Basic materials$^8$ \\
11 & OXY & Energy$^1$ & 56 & SKY & Capital goods & 101 & AA & Basic materials$^8$ \\
12 & RD & Energy$^1$ & 57 & FLE & Capital goods & 102 & PD & Basic materials$^8$ \\
13 & MRO & Energy$^1$ & 58 & RSH & Services & 103 & LPX & Basic materials$^8$ \\
14 & XOM & Energy$^1$ & 59 & EK & Consumer cyclical & 104 & N & Basic materials$^8$ \\
15 & PGL & Utilities$^2$ & 60 & EMR & Conglomerates & 105 & DE & Capital goods \\
16 & CNP & Utilities$^2$ & 61 & TOY & Services & 106 & PBI & Technology \\
17 & ETR & Utilities$^2$ & 62 & TEN & Consumer cyclical & 107 & BDK & Consumer cyclical \\
18 & DTE & Utilities$^2$ & 63 & ROK & Technology & 108 & UNP & Transportation$^9$ \\
19 & EXC & Utilities$^2$ & 64 & HON & Capital goods & 109 & NSC & Transportation$^9$ \\
20 & AEP & Utilities$^2$ & 65 & AXP & Financial & 110 & CSX & Transportation$^9$ \\
21 & PEG & Utilities$^2$ & 66 & GRA & Basic materials & 111 & BNI & Transportation$^9$ \\
22 & SO & Utilities$^2$ & 67 & VVI & Services & 112 & CNF & Transportation$^9$ \\
23 & ED & Utilities$^2$ & 68 & CSC & Technology$^6$ & 113 & MAT & Consumer cyclical \\
24 & PCG & Utilities$^2$ & 69 & DBD & Technology$^6$ & 114 & C & Financial \\
25 & EIX & Utilities$^2$ & 70 & HRS & Technology$^6$ & 115 & VIA & Services \\
26 & LMT & Capital goods$^3$ & 71 & STK & Technology$^6$ & 116 & MMM & Conglomerates \\
27 & NOC & Capital goods$^3$ & 72 & ZL & Technology$^6$ & 117 & DIS & Services \\
28 & RTN & Conglomerates$^3$ & 73 & TEK & Technology$^6$ & 118 & BC & Consumer cyclical \\
29 & GD & Capital goods$^3$ & 74 & AVT & Technology$^6$ & 119 & CBE & Technology \\
30 & BA & Capital goods$^3$ & 75 & GLW & Technology$^6$ & 120 & THC & Healthcare$^{10}$ \\
31 & BOL & Healthcare$^4$ & 76 & NSM & Technology$^6$ & 121 & HUM & Financial$^{10}$ \\
32 & MDT & Healthcare$^4$ & 77 & TXN & Technology$^6$ & 122 & AET & Financial$^{10}$ \\
33 & BAX & Healthcare$^4$ & 78 & MOT & Technology$^6$ & 123 & CI & Financial$^{10}$ \\
34 & WYE & Healthcare$^4$ & 79 & HPQ & Technology$^6$ & 124 & JCP & Services \\
35 & BMY & Healthcare$^4$ & 80 & NT & Technology$^6$ & 125 & MEE & Energy \\
36 & LLY & Healthcare$^4$ & 81 & IBM & Technology$^6$ & 126 & GE & Conglomerates \\
37 & MRK & Healthcare$^4$ & 82 & UIS & Technology$^6$ & 127 & UTX & Conglomerates \\
38 & PFE & Healthcare$^4$ & 83 & XRX & Technology$^6$ & 128 & R & Services \\
39 & JNJ & Healthcare$^4$ & 84 & T & Services & 129 & NVO & Healthcare \\
40 & PEP & Consumer noncyclical$^5$ & 85 & HIT & Capital goods & 130 & GT & Consumer cyclical \\
41 & KO & Consumer noncyclical$^5$ & 86 & MER & Financial & 131 & S & Services \\
42 & PG & Consumer noncyclical$^5$ & 87 & FDX & Transportation$^7$ & 132 & NAV & Consumer cyclical \\
43 & MO & Consumer noncyclical$^5$ & 88 & LUV & Transportation$^7$ & 133 & CEN & Technology \\
44 & CL & Consumer noncyclical$^5$ & 89 & DAL & Transportation$^7$ & 134 & FL & Services
\end{tabular}
\end{ruledtabular}
\end{table*}

Figure~\ref{fig:block} visualizes the correlation
matrix elements $C^g_{l_i l_j}$ with the most optimized sequence $\{l_i\}$
and Table~\ref{tab} lists the optimized sequence of stocks.
The multiple independent blocks of highly correlated correlations
in the matrix are clearly visible without any \textit{a priori} knowledge of
stocks, i.e., the stocks in different blocks are believed to
belong to different groups.
We succeed to identify about $70\%$ of the entire $135$ stocks from the
blocks, which are listed in Table~\ref{tab} and
it turns out that most of the stocks in a block are represented by
a single industry sector or a detailed industrial classification
such as aerospace and defense, airline transport,
railroad, and insurance (see Fig.~\ref{fig:block}).
There still remain a small number of ungrouped stocks,
which arises from the fact
that the correlations between them are too weak to be
distinguished from the random noise that still exists in
the group correlation matrix.

As an alternative method,
we also perform a network-based approach to find the groups of stocks.
In principle, the correlation matrix can be treated as an
adjacency matrix of the weighted network of stocks, in which
the weights indicate how closely correlated the stocks are in their
price changes \cite{Swson}.
However, for the simplicity and the clear definition of groups in the network,
we consider the binary network of stocks which permits only two possible
states of a stock pair, connected or disconnected.

To construct the binary network of stocks, we use the percolation approach
because of its usefulness of finding groups. The method is very simple:
for each pair of stocks, we connect them if the group correlation coefficient
$C^g_{ij}$ is larger than a preassigned threshold value $p$.
If the heavy tail in the distribution of $C^g_{ij}$
in Fig.~\ref{fig:distribution} mostly comes from the correlation
between the stocks in the same group, an appropriate choice of $p=p_c$
will give several meaningful isolated clusters, $m$, in the network
which are expected to be identified as different stock groups.

\begin{figure*}
\begin{center}
\includegraphics[width=0.8\textwidth]{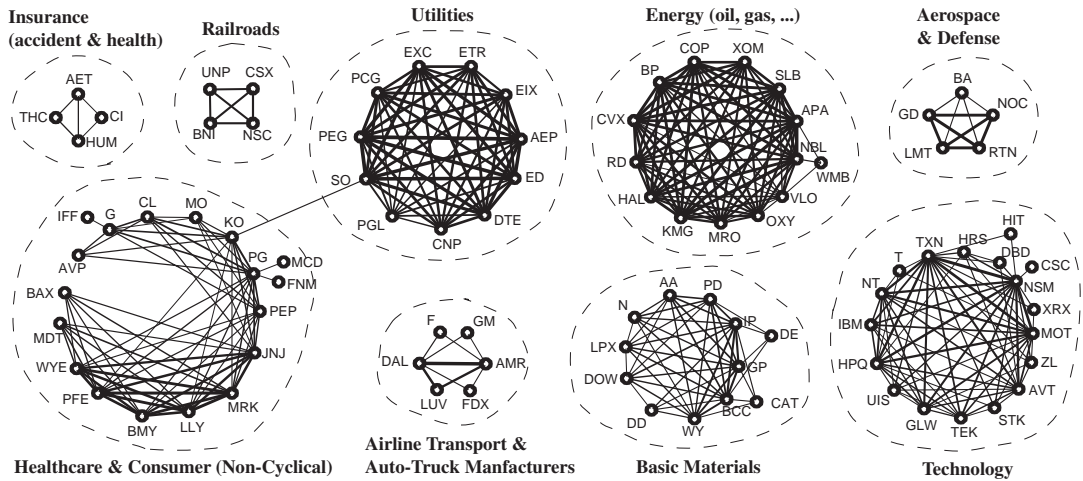}
\caption{
\label{fig:network}
The stock network with the threshold $p_c=0.1$. The thickness of
links indicates the strength of the correlation in the group correlation
matrix.
}
\end{center}
\end{figure*}

We determine $p_c$ by observing the change of the network
structure as $p$ decreases. Figure~\ref{fig:cluster}(a) displays
the number of isolated clusters in the network as a function of threshold
$p$. As we decrease $p$, the number of isolated clusters in the network
increases slowly and stays near the maximum value up to $p=0.1$, and then
it abruptly decreases to $1$, which indicates there exists only one isolated
cluster. Therefore we choose $p_c=0.1$ to construct the most
clustered but stable stock network~\cite{comment3}.

We find that the constructed network consists of separable
groups of stocks which correspond to the industrial sectors of stocks
(see Fig.~\ref{fig:network}). At $p_c=0.1$, the network has $92$
nodes and $357$ links. The identification of stock group
is very clear because the clusters in the network, which
we consider to be equivalent to stock groups, are fully connected
networks or very dense networks
in which most of the nodes in the cluster are directly connected.
However, although most of the stock groups are represented by a
single industrial sector, it is found that the stocks
which belong to two different industrial sectors coexist in a cluster.
For instance, the stocks in the healthcare sector and
the noncyclical consumer sector cannot be separable in this network.
Indeed, in Fig.~\ref{fig:block}, one can observe non-negligible
correlation between the healthcare and the noncyclical consumer,
which indicates the presence of an intergroup correlation.
In the real market, this presence of such an intersector correlation 
can be expected and our clustering results shown 
in Figs.~\ref{fig:block} and \ref{fig:network} present 
both of intergroup and intragroup correlations 
that exist in the real stock market. 

The group identification based on the eigenvector analysis
of the stock price correlation matrix has been studied by several
research groups~\cite{Plerou2,Gopikrishnan,Utsugi}.
In spite of their pioneering achievements to reveal the localization
properties of eigenvectors, the classification of stocks into groups
was not so clear, and it only covered about $10\%$ of their stocks
because they used only the few highest contributions of eigenvector components
due to the ambiguity explained in Sec.~\ref{sec:intro}.
In this work, we not only introduce a more refined and systematic method to
identify the stock groups, but also successfully cluster about $70\%$
of stocks into groups although direct comparison of
the success ratio might be inappropriate because our data set is
different from theirs.

On the other hand, Onnela {\it et al.}~\cite{Onnela2}
introduced the percolation approach
to construct the stock network in which the links are added between stocks
one by one in descending order from the highest element of the
full correlation matrix. In their work, though highly correlated groups
of stocks were found, the threshold value of the correlation
to settle the network structure was hardly determined; the number of
isolated clusters according to the threshold did not show
the clear cut. We believe that this is attributed to the fact that
they used the full correlation matrix carrying marketwide and
random fluctuation. We would also fail to determine the critical threshold value
of correlation if we use the full correlation matrix instead of the filtered one
[see Fig.~\ref{fig:cluster}(b)]. This indicates that the filtering
is crucial for the stock group identification.

Finally, we note that Marsili {\it et al.} introduced a different
method to filter noises from the time series of stock price
log-returns for stock group identification. In their work, it was
assumed that the normalized log-return could be expressed by the
linear combination of the noise at individual stock level and the
noise at the level of the groups, which fitted to the real data to
determine the weights of two noises and the constituents of the
groups. However, we found that the effect of the inhomogeneous 
marketwide fluctuation is quite significant that the marketwide effect 
needs to be considered seriously to describe the correlation between 
stock correctly. Indeed, it is found that the filtering out of the
corresponding $\mathrm{C}^m$ improves the clustering result.

\section{Conclusion}
\label{sec:conclusion}

In conclusion, we successfully identify the multiple group of stocks
from the empirical correlation matrix of stock price changes in the New York
Stock Exchange. We propose refined methods to find stock groups which
dramatically reduce ambiguities as compared to identifying stock groups
from the localization in a single eigenvector of the correlation matrix
\cite{Plerou2,Gopikrishnan,Utsugi}.
From the analysis of the characteristics of eigenvectors, we construct
the group correlation matrix of the stock groups excluding
the marketwide effect and random noise. By optimizing the representation
of the group correlation matrix, we find that the group correlation
matrix is represented by the block diagonal matrix where
the stocks in each block belong to the same group. This coincides
with the theoretical model of Noh~\cite{Noh}. Equally good stock
group identification is also achieved by the percolation approach
on the group correlation matrix to construct the network of stocks.

\acknowledgments
We thank J. D. Noh for helpful discussions.
This work was supported by
grant No. R14-2002-059-01002-0 from KOSEF-ABRL program.


\begin{thebibliography}{}
\bibitem{comment1} The similar matrix element rearranging technique was
used in Ref.~\cite{Bilke}.
\bibitem{comment2} The visualization of $\mathbf{C}^g$ is not very
sensitive to the cutoff value $c_c$.
\bibitem{comment3} Small change of $p_c$, $\sim 0.01$, does not
seriously change the clustering result.
\end{thebibliography}
\end{document}